\documentclass[nofootinbib,preprint,superscriptaddress, pra]{revtex4-1}
\usepackage{amsfonts}
\usepackage[dvips]{color}
\usepackage{newlfont}
\usepackage{graphicx}
\usepackage{amsmath}
\usepackage[latin1]{inputenc}
\usepackage{float}
\usepackage{times}
\usepackage{subfig}

\begin{document}

\author{ L. F. Lopes Oliveira}
\affiliation{Programa de Pós-Graduação em Modelagem Matemática e Computacional,
Centro Federal de Educa{çã}o Tecnológica de Minas Gerais, 30510-000, Belo Horizonte, MG, Brazil}
\author{R. Rossi, Jr.}
\affiliation{Universidade Federal de Viçosa, Campus Florestal, 35690-000 - Florestal, MG - Brasil}
\author{A. R. Bosco de Magalh{ã}es}
\author{J. G. Peixoto de Faria}
\affiliation{Programa de Pós-Graduação em Modelagem Matemática e Computacional,
Centro Federal de Educa{çã}o Tecnológica de Minas Gerais, 30510-000, Belo Horizonte, MG, Brazil}
\affiliation{Departamento de F{í}sica e Matemática, Centro Federal de Educa{çã}o Tecnoló%
gica de Minas Gerais, 30510-000, Belo Horizonte, MG, Brazil}
\author{M. C. Nemes}
\affiliation{
Departamento de Física, Instituto de Ciências Exatas, Universidade Federal de Minas Gerais,
CP 702, 30161-970, Belo Horizonte, MG, Brazil}

\title{Continuous Monitoring of Dynamical Systems and Master Equations}

\begin{abstract}
We illustrate the equivalence between the non-unitary evolution
of an open quantum system governed by a Markovian master equation
and a process of continuous measurements involving this system.
We investigate a system of two coupled modes, only one of them
interacting with external degrees of freedom, represented,
in the first case, by a finite number of harmonic oscillators, and,
in the second, by a sequence of atoms where each one interacts with a single
mode during a limited time. Two distinct regimes appear, one of them corresponding
to a Zeno-like behavior in the limit of large dissipation.
\end{abstract}

\maketitle

The decoherence program is one of the most successfull theroretical
proposals devised to give suitable responses to the problem of quantum-to-classical transition
(see \cite{schlosshauer2005} and references therein) within the scope of the quantum
theory itself. According to this program,
isolated systems must be considered as mere idealization and the emergence of classicality
from quantum mechanics is a consequence of the unavoidable interaction between a
physical system and its environment. The net effect of this interaction is
the disappearance of the local quantum correlations between preferred states\footnote{To be exact,
the quantum correlations does not disappear. In fact, they are carried from the system
to the environment and turned out inaccessible.}. The early formulations of this program
assign the mechanism of decoherence to the continuous monitoring of the
system of interest by the surrounding particles \cite{simonius1978, harris1981, harris1982, art1}.
These formulations were devised to explain the quasilocalization of macroscopic bodies or symmetry breaking
in microscopic systems, like the
observation of a well defined chirality in optical isomers.

One of the widespread ways to obtain the \textit{effective} dynamics of an open quantum system
considers the microscopic model describing the coupling between the system
of interest and the degrees of freedom of its environment.
In general, the number of these environmental degrees of freedom considered  is
very large, such that the irreversibility of the effective dynamics could be guaranteed.
In the limit of infinite external degrees of freedom, the environment is
suitably modeled by a thermal reservoir and the
effective dynamics of the system of interest is governed by a master equation obtained by tracing
these external degrees of freedom out \cite{kubo}.

Another well known effect related to continuous
measurements is the Quantum Zeno Effect (QZE). In 1977 B. Misra and E. C. G. Sudarshan
reported an intriguing result related to measurements in Quantum Mechanics \cite{art2}.
They showed that a sequence of projective measurements on a system inhibits its time evolution.
In the limit of continuous measurements the evolution is completely frozen. Similarities with
one of the paradoxes proposed by Zeno of Elea, who intend to show that movement is theoretically
impossible, motivated the authors to name the quantum effect after the Greek philosopher.
Originally the Quantum Zeno Effect was called the Quantum Zeno Paradox, because it was
supposed to show the theoretical impossibility of a ``movement" (the quantum evolution)
of a decaying particle in a bubble chamber.

The Quantum Zeno Effect (QZE)
plays an important role in Quantum Mechanics and enormous
literature about this topic has been produced over the last thirty years.
After the realization of the pioneer
experiment \cite{art4} on the effect, which showed the inhibition of
transitions between quantum states by means of frequent observations, the QZE became
the center of fervorous debates
\cite{art5,art6} about the necessity of the projection postulate on the measurement description.
New approaches have been
proposed \cite{art5,art7} and the initial association between the QZE and the projection
postulate was no longer a necessary ingredient. Nowadays the literature on this subject
is extensive and includes relation between QZE and quantum jumps \cite{art8,art9}, quantum
Zeno dynamics \cite{art10,art11}, its implementation in the system of microwave cavities
\cite{art12}, and semiclassical evolution for coupled systems obtained by frequent Zeno-like
measurements \cite{art13}. With the increasing interest in quantum computation, QZE has become
also a tool for the development of protocols on quantum state protection \cite{art14,art15,art16},
that are important for the implementation of quantum computation.

Although originally the QZE was considered as a result of a sequence of measurements on the system of interest,
the effect has also been predicted in different contexts, as an example we quote: In Ref.\cite{har1}
the authors suggested that a manifestation of the QZE appears in the relaxation of optical activity
in a medium. In Ref.\cite{har2} a relation between the QZE and the coherence properties of a two-state system
subjected to random influences in a medium is presented. In Ref.\cite{art17,art18} the authors also show that
the QZE can be induced by other physical interactions.

In the present work we study two completely distinct approaches to the
dynamics of quantum open systems.
We consider the system composed by two coupled modes interacting unitarily,
where only one of them is coupled to
external degrees of freedom.
We show that the Zeno like effect
similar to the one shown in \cite{art18} is present in both dynamics. Firstly we couple to the
system of interest a dynamical environment over which we have control (interactions, number of
components, etc.). Secondly we simulate an experimental situation where the system is continuously
monitored by a probe system, again controlling interaction time, and other parameters.

We show that not only the conjecture is correct in the situations presented,
but also a Zeno like effect can be obtained for both cases, when the
influence of the external system is sufficiently strong. Precisely the same effect has been observed by
modeling the tunneling of a photon between two cavities, one of them whose
dissipation is governed by a master equation \cite{art18}.

\section{Coupling with a finite number of harmonic oscillators}

In this section we consider the system of two linearly coupled harmonic oscillators, one of them
coupled to an environment composed by a finite number of harmonic
oscillators according to the Hamiltonian
\begin{eqnarray}
\hat{H} &=&\hat{H}_{S}+\hat{H}_{R}+\hat{H}_{int}\text{,\qquad }\hat{H}%
_{S}=\omega _{1}\hat{a}_{1}^{\dagger }\hat{a}_{1}+\omega _{2}^{\dagger }\hat{%
a}_{2}^{\dagger }\hat{a}_{2}+G\left( \hat{a}_{1}^{\dagger }\hat{a}_{2}+\hat{a%
}_{2}^{\dagger }\hat{a}_{1}\right) ,  \label{H1} \\
\hat{H}_{R} &=&\overset{N}{\underset{k=3}{\sum }}\omega _{k}^{\dagger }\hat{a%
}_{k}^{\dagger }\hat{a}_{k}\text{,\qquad }\hat{H}_{int}=\overset{N}{%
\underset{k=3}{\sum }}\gamma _{k}\left( \hat{a}_{1}^{\dagger }\hat{a}_{k}+%
\hat{a}_{k}^{\dagger }\hat{a}_{1}\right) ,  \notag
\end{eqnarray}
where $\hat{a}_{1}^{\dagger }$ and $\hat{a}_{2}^{\dagger }$ ($\hat{a}_{1}$
and $\hat{a}_{2}$) are creation (annihilation) bosonic operators for the
modes of interest $M_{1}$ and $M_{2}$, and $\hat{a}_{k}^{\dagger }$ and $%
\hat{a}_{k}$ refer to the environment for $k$ ranging from $k=3$ to $N$. Defining
\begin{equation}
\mathbf{A}=\left(
\begin{array}{c}
\hat{a}_{1} \\
\hat{a}_{2} \\
\hat{a}_{3} \\
\vdots \\
\hat{a}_{N}
\end{array}
\right) \text{,\qquad }\mathbf{A}_{\mathbf{D}}=\left(
\begin{array}{c}
\hat{a}_{1}^{\dagger } \\
\hat{a}_{2}^{\dagger } \\
\hat{a}_{3}^{\dagger } \\
\vdots \\
\hat{a}_{N}^{\dagger }
\end{array}
\right) \text{,\qquad }\mathbf{H}=\left(
\begin{array}{ccccc}
\omega _{1} & G & \gamma _{3} & \cdots & \gamma _{N} \\
G & \omega _{2} & 0 & \cdots & 0 \\
\gamma _{3} & 0 & \omega _{3} & \cdots & 0 \\
\vdots & \vdots & \vdots & \ddots & \vdots \\
\gamma _{N} & 0 & 0 & \cdots & \omega _{N}
\end{array}
\right) \text{,}
\end{equation}
we can write the Hamiltonian $\hat{H}$ in a matrix form:
\begin{equation}
\hat{H}=\mathbf{A}_{\mathbf{D}}^{T}\mathbf{HA}.
\end{equation}
In order to preserve the hermiticity of the operator $\hat{H}$, the matrix $\mathbf{H}$
must be hermitian. Without loss of generality, we can consider it real and symmetric. Thus, there is an orthogonal matrix $%
\mathbf{P}$ such that
\begin{equation}
\mathbf{P}^{T}\mathbf{HP}=\left(
\begin{array}{cccccc}
\lambda _{1} & 0 & \cdots & 0 \\
0 & \lambda _{2} & \cdots & 0 \\
\vdots & \vdots & \ddots & \vdots \\
0 & 0 & \cdots & \lambda _{N}
\end{array}
\right) =\mathbf{D}\text{,}
\end{equation}where the $\lambda _{k}$ are real numbers, which may be used to write the
Hamiltonian in a diagonal form:
\begin{equation}
\hat{H}=\mathbf{B}_{\mathbf{D}}^{T}\mathbf{DB=}\overset{M}{\underset{k=1}{%
\sum }}\lambda _{k}\hat{b}_{k}^{\dagger }\hat{b}_{k}\text{,}
\end{equation}
where
\begin{equation}
\mathbf{B}=\mathbf{P}^{T}\mathbf{A=}\left(
\begin{array}{c}
\hat{b}_{1} \\
\hat{b}_{2} \\
\vdots \\
\hat{b}_{N}
\end{array}
\right) \text{,\qquad }\mathbf{B}_{\mathbf{D}}=\mathbf{P}^{T}\mathbf{A}_{%
\mathbf{D}}=\left(
\begin{array}{c}
\hat{b}_{1}^{\dagger } \\
\hat{b}_{2}^{\dagger } \\
\vdots \\
\hat{b}_{N}^{\dagger }
\end{array}
\right) \text{.}  \label{B=PTA}
\end{equation}
Using the orthogonality of $\mathbf{P}$, it is easy to show that canonical
commutation relations hold for $\hat{b}_{k}$ and $\hat{b}_{k}^{\dagger }$:
\begin{equation}
\left[ \hat{b}_{k},\hat{b}_{j}^{\dagger }\right] =\delta _{kj}\text{,\qquad }%
\left[ \hat{b}_{k},\hat{b}_{j}\right] =\left[ \hat{b}_{k}^{\dagger },\hat{b}%
_{j}^{\dagger }\right] =0\text{.}
\end{equation}
The modes related to the operators $\hat{a}_{k}$ shall be called the
original modes; the ones concerning to $\hat{b}_{k}$ will be referred to as the
normal modes.

In what follows,
we assume that the main modes $M_1$ and $M_2$ are resonant and we consider that the matrix $\mathbf{H}$ is a
function of five positive parameters $N$, $\Omega $, $G$, $\Delta $ and $%
\Gamma $:
\begin{equation}
\mathbf{H}=\left(
\begin{array}{ccccccccc}
\Omega & G & \frac{\Gamma }{\sqrt{N-2}} & \frac{\Gamma }{\sqrt{N-2}} & \frac{%
\Gamma }{\sqrt{N-2}} & \frac{\Gamma }{\sqrt{N-2}} & \cdots & \frac{\Gamma }{%
\sqrt{N-2}} & \frac{\Gamma }{\sqrt{N-2}} \\
G & \Omega & 0 & 0 & 0 & 0 & \cdots & 0 & 0 \\
\frac{\Gamma }{\sqrt{N-2}} & 0 & \Omega +\frac{\Delta }{N-2} & 0 & 0 & 0 &
\cdots & 0 & 0 \\
\frac{\Gamma }{\sqrt{N-2}} & 0 & 0 & \Omega -\frac{\Delta }{N-2} & 0 & 0 &
\cdots & 0 & 0 \\
\frac{\Gamma }{\sqrt{N-2}} & 0 & 0 & 0 & \Omega +2\frac{\Delta }{N-2} & 0 &
\cdots & 0 & 0 \\
\frac{\Gamma }{\sqrt{N-2}} & 0 & 0 & 0 & 0 & \Omega -2\frac{\Delta }{N-2} &
\cdots & 0 & 0 \\
\vdots & \vdots & \vdots & \vdots & \vdots & \vdots & \ddots & \vdots &
\vdots \\
\frac{\Gamma }{\sqrt{N-2}} & 0 & 0 & 0 & 0 & 0 & \cdots & \Omega +\frac{%
\Delta }{2} & 0 \\
\frac{\Gamma }{\sqrt{N-2}} & 0 & 0 & 0 & 0 & 0 & \cdots & 0 & \Omega -\frac{%
\Delta }{2}
\end{array}
\right) \text{.}  \label{H2}
\end{equation}
The parameter $\Omega$ gives the frequency of the oscillators of interest,
which are coupled according to the constant $G$.
The total number of environmental oscillators is ($N-2$).
Their frequencies are distributed in a finite interval around $\Omega$ defined with the help of the parameter $\Delta$:
$\left[ \Omega -\Delta /2, \Omega + \Delta /2 \right]$.
If $\Delta$ is large enough, the environmental modes out of this interval have negligible effects over the oscillators of interest \cite{art19}.
The environmental frequencies are equally spaced distributed above and below the central frequency $\Omega$.
This choice is not essential for the dynamics induced and for derivation of master equations:
in fact, randomly distributed frequencies lead, when $N$ is large, to equivalent results.
In derivations of master equations, it is usually assumed reservoirs with infinite frequency range and dense spectrum.
Infinite frequency range is approximated by increasing $\Delta$, and dense spectrum by decreasing
$\Delta / \left( N-2 \right)$;
in order to achieve both conditions simultaneously, we must have large values of $\Delta$ and $N$.
In this limit, a master equation may derived from the Hamiltonian specified by Eq. (\ref{H2}).
The parameter $\Gamma$ is related to the strength of the
coupling with the environment.
The choice of $\gamma _{k}$ being proportional to $1 / \sqrt{N-2}$ is consistent with assumptions usually
performed in derivation of master equations that guarantee finite decay
rates in the thermodynamic limit $N\longrightarrow \infty $.
Of course, all $\gamma _{k}$ do not have to be equal for such derivations.

Since the $\hat{b}_{k}$ are linear combinations of the $\hat{a}_{k}$, they
share the same vacuum state $\left| 0\right\rangle $. In order to
investigate the dynamics of the system plus environment in the space of one
excitation, let us define the states
\begin{equation}
\left| \theta _{k}\right\rangle =\hat{a}_{k}^{\dagger }\left| 0\right\rangle
\text{,\qquad }\left| \phi _{k}\right\rangle =\hat{b}_{k}^{\dagger }\left|
0\right\rangle \text{,\qquad }k=1\text{ to }N\text{.}
\end{equation}
Using Eqs. (\ref{B=PTA}), we see that they are connected by
\begin{equation}
\left| \theta _{k}\right\rangle =\overset{N}{\underset{j=1}{\sum }}%
P_{k,j}\left| \phi _{j}\right\rangle \text{,\qquad }\left| \phi
_{k}\right\rangle =\overset{N}{\underset{j=1}{\sum }}P_{j,k}\left| \theta
_{j}\right\rangle \text{,}
\end{equation}
where $P_{i,j}$ is the element of the matrix $\mathbf{P}$ in the $i$-th
line and $j$-th column. Using these relations and observing that $\hat{H}%
\left| \phi _{k}\right\rangle =\lambda _{k}\left| \phi _{k}\right\rangle $,
we can calculate the evolution of the states with one excitation in the
original modes as
\begin{equation}
e^{-i\hat{H}t}\left| \theta _{k}\right\rangle =\overset{N}{\underset{j=1}{%
\sum }}P_{k,j}e^{-i\hat{H}t}\left| \phi _{j}\right\rangle =\overset{N}{%
\underset{l=1}{\sum }}\left( \overset{N}{\underset{j=1}{\sum }}%
P_{k,j}P_{l,j}e^{-i\lambda _{j}t}\right) \left| \theta _{l}\right\rangle
\text{.}
\end{equation}

\begin{figure}[h]
	\subfloat[]{\label{fig1a}\includegraphics[scale=0.5]{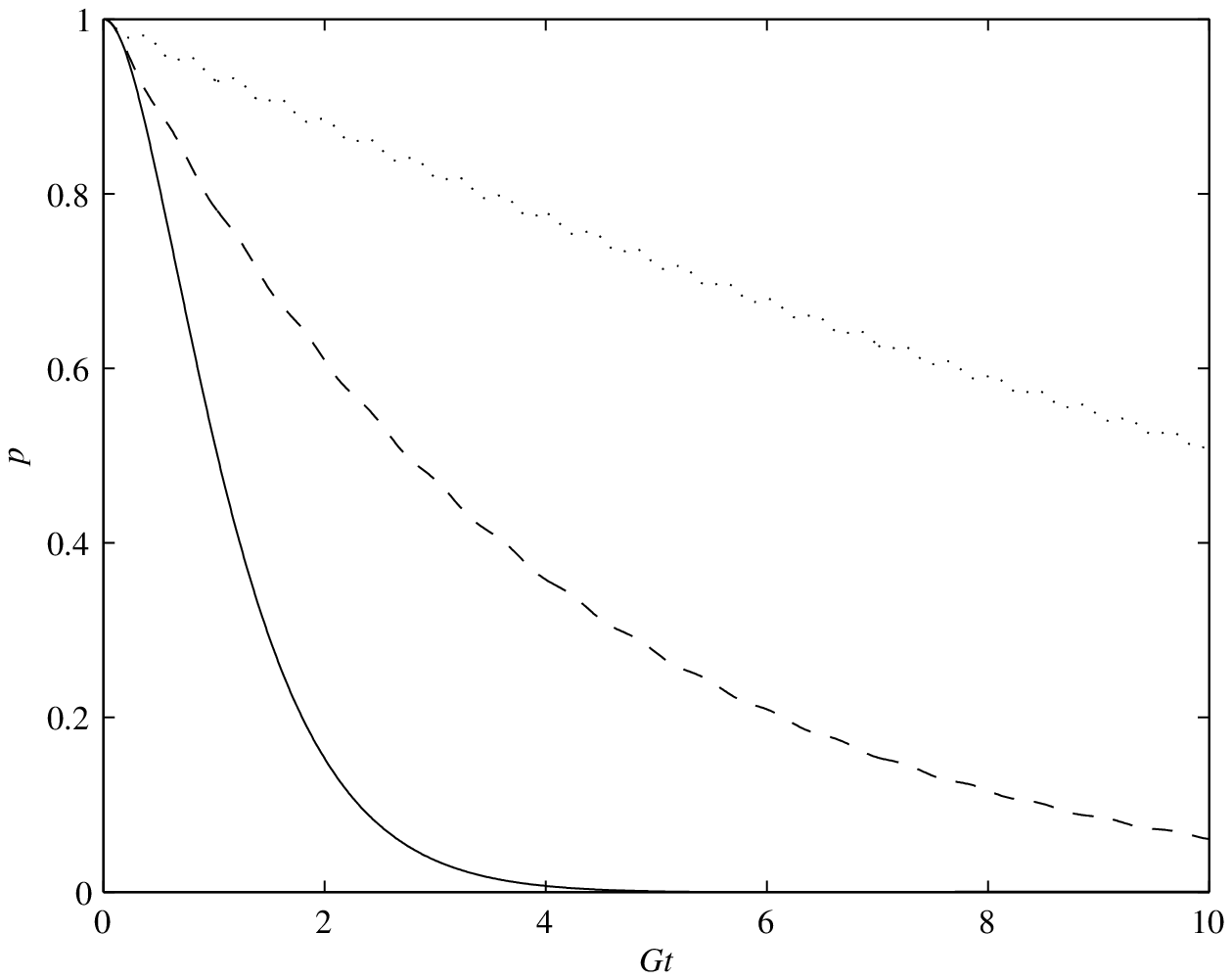}}
	\subfloat[]{\label{fig1b}\includegraphics[scale=0.5]{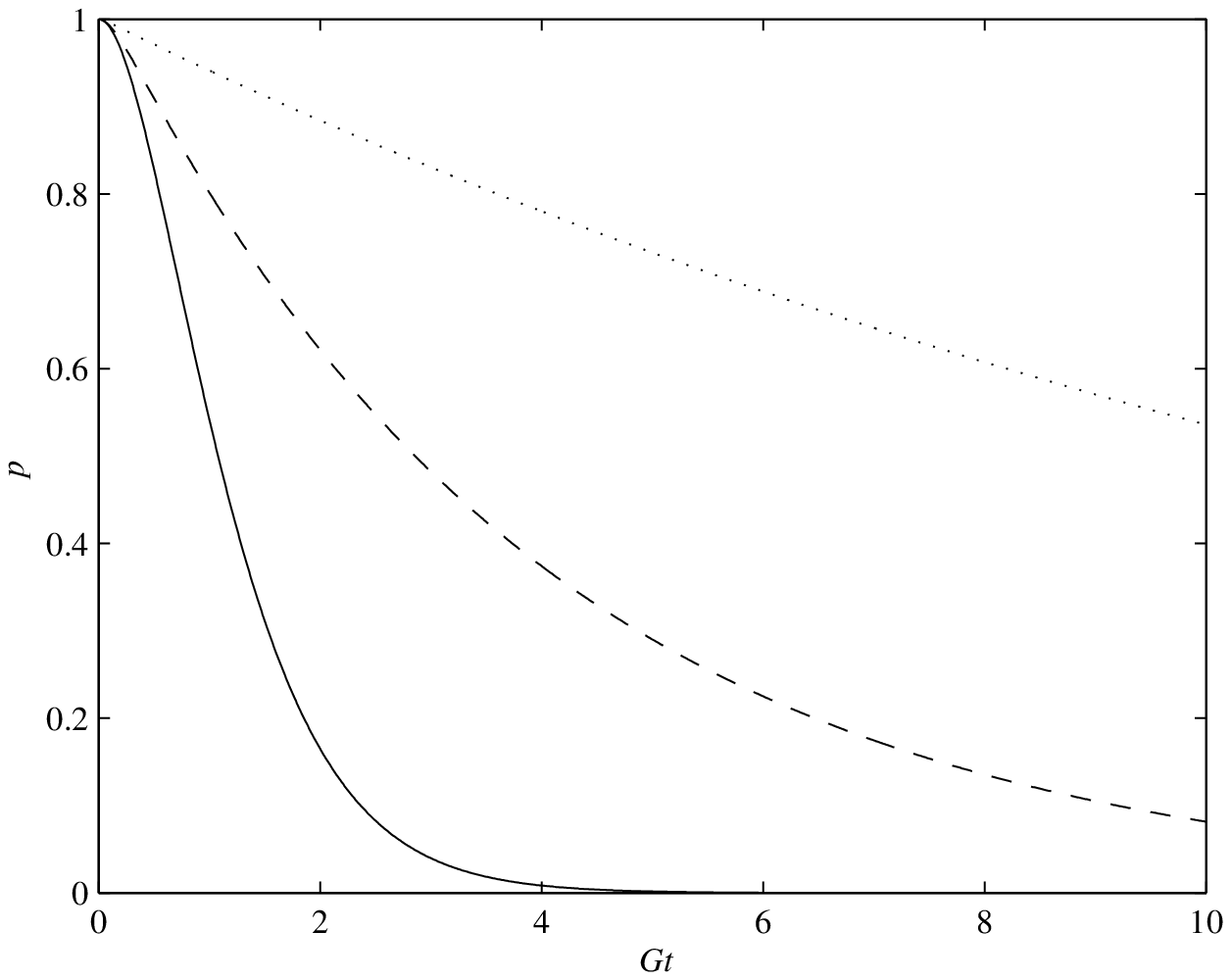}}\\
	\subfloat[]{\label{fig1c}\includegraphics[scale=0.21]{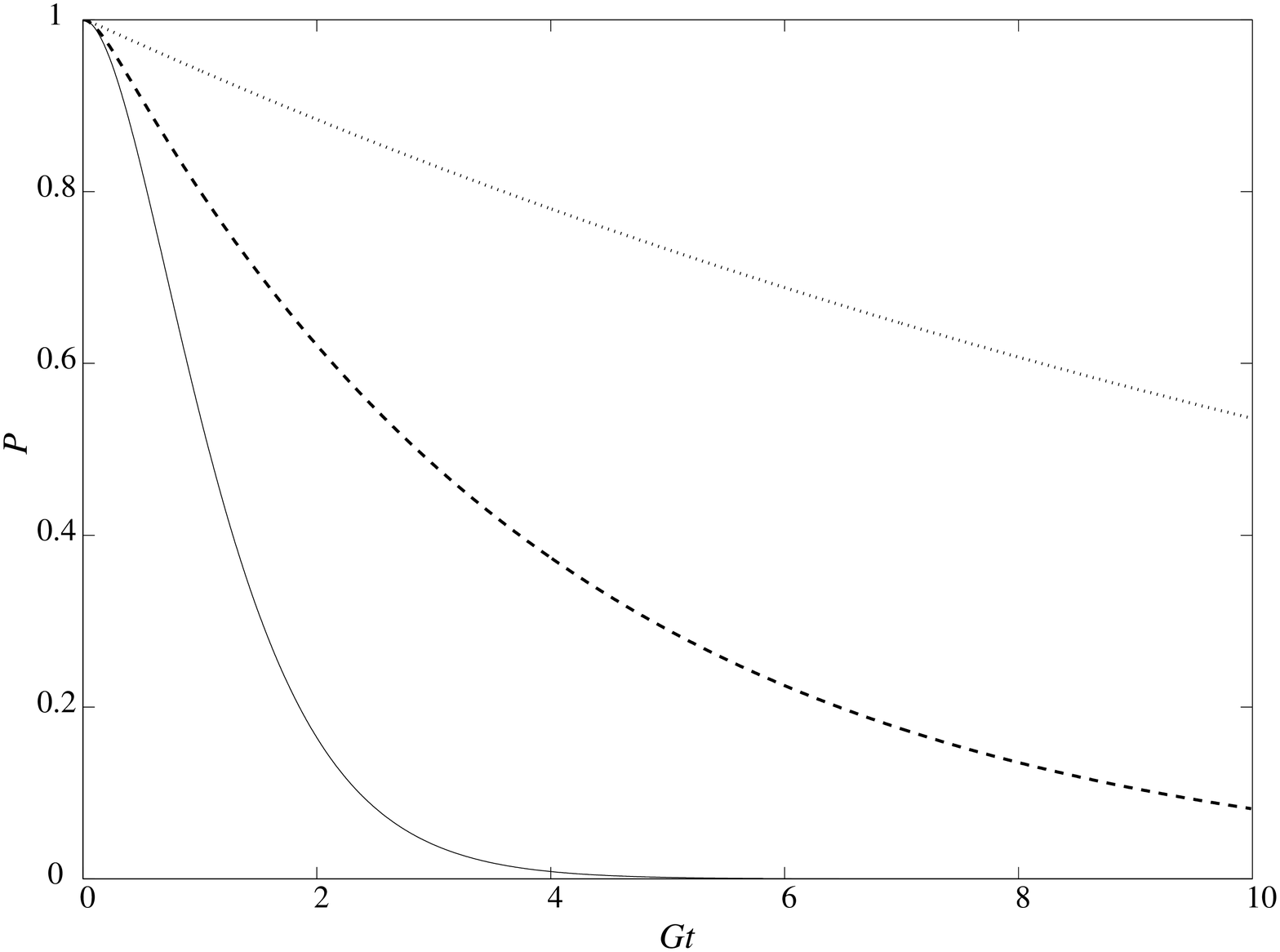}}
	\label{zenoregime}
	\caption{(a) Evolution of the occupation of mode $M_{2}$ according to the Hamiltonian approach
	for $N=500$, $\Omega=100G$, $\Delta=20G$ and $\Gamma=\sqrt{\frac{\eta\Delta G}{\pi }}$
	with $\eta=2$ (solid line), $\eta=8$ (dashed line) and $\eta=32$ (dotted line).
	(b) Evolution of
	the occupation of mode $M_{2}$ according to the master equation approach
	for $\kappa/G=2$ (solid line), $\kappa/G=8$ (dashed line) and $\kappa/G=32$
	(dotted line). Figure from Ref.	\cite{art18}. (c) Evolution of the occupation of
	mode $M_{2}$ during a sequence of measurements probing the presence of the excitation
	in mode $M_1$ for $g=\sqrt{\frac{2\eta G}{t_{int}}}$ with $\eta = 2$ (solid line),
	$\eta = 8$ (dashed line) and $\eta = 32$ (dotted line).
	For sake of comparison, in the plots (a) and (c), a curve concerning a given value of $\eta$
	corresponds to the curve related to the same value of $\kappa/G$ in plot (b).}
\end{figure}

\begin{figure}[h]
	\subfloat[]{\label{fig2a}\includegraphics[scale=0.5]{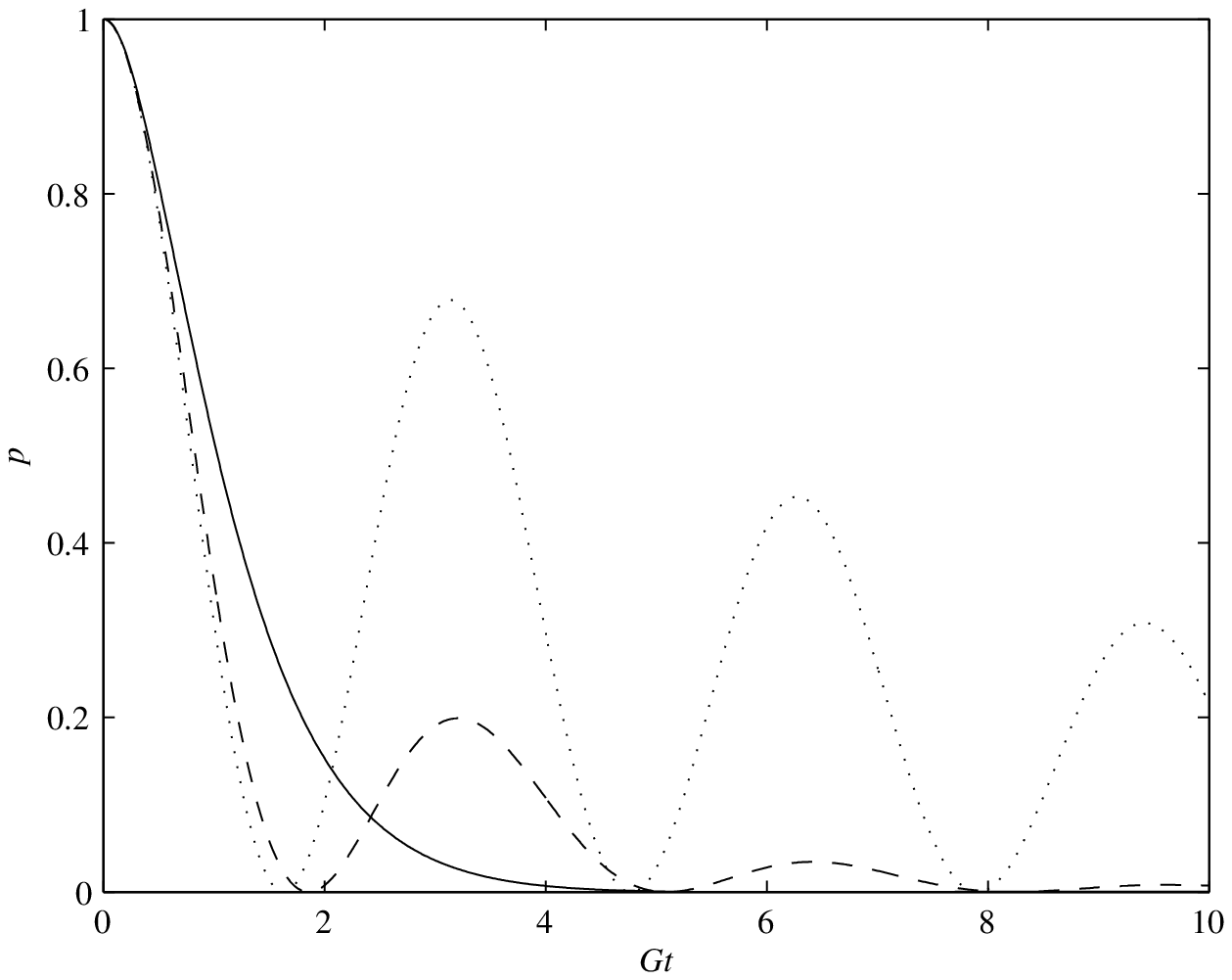}}
	\subfloat[]{\label{fig2b}\includegraphics[scale=0.5]{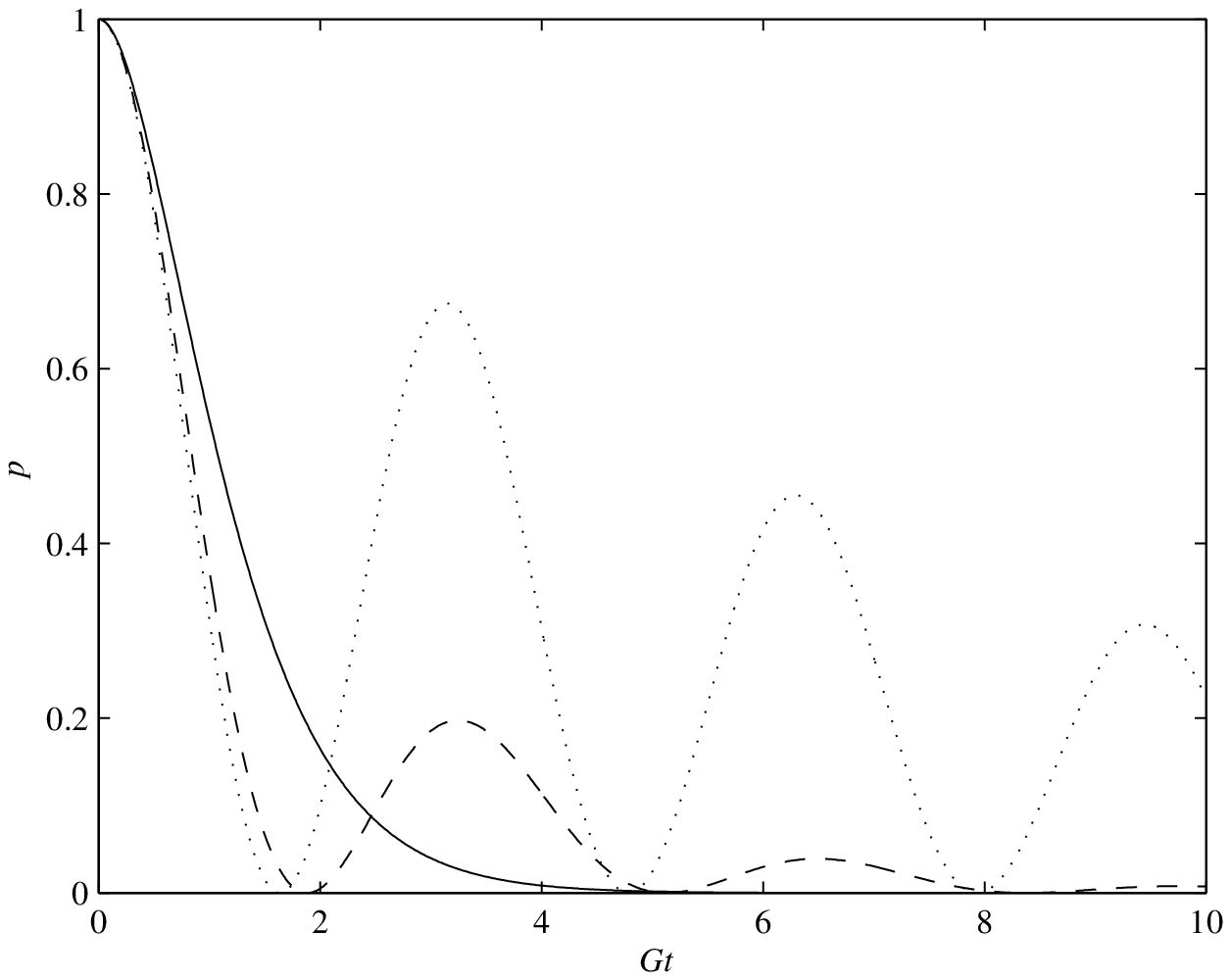}}\\
	\subfloat[]{\label{fig2c}\includegraphics[scale=0.21]{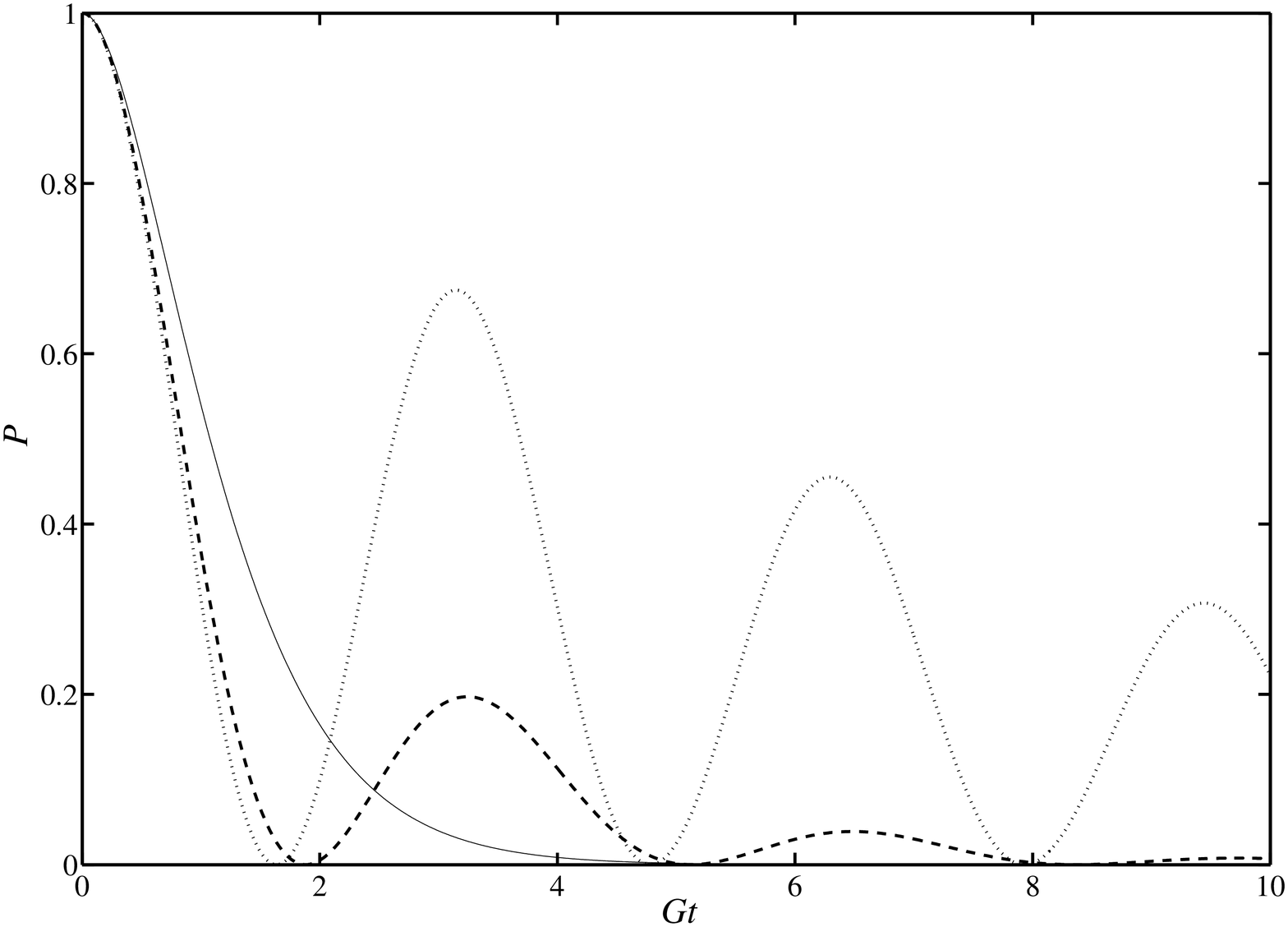}}
	\label{nonzenoregime}
	\caption{(a) Evolution of the occupation of mode $M_{2}$ according to the Hamiltonian approach
	for $N=500$, $\Omega=100G$, $\Delta=20G$ and $\Gamma=\sqrt{\frac{\eta\Delta G}{\pi }}$
	with $\eta=2$ (solid line), $\eta=1/2$ (dashed line) and $\eta=1/8$ (dotted line).
	(b) Evolution of
	the occupation of mode $M_{2}$ according to the master equation approach
	for $\kappa/G=2$ (solid line), $\kappa/G=1/2$ (dashed line) and $\kappa/G=1/8$
	(dotted line). Figure from Ref.	\cite{art18}. (c) Evolution of the occupation of
	mode $M_{2}$ during a sequence of measurements probing the presence of the excitation
	in mode $M_1$ for $g=\sqrt{\frac{2\eta G}{t_{int}}}$ with $\eta = 2$ (solid line),
	$\eta = 1/2$ (dashed line) and $\eta = 1/8$ (dotted line).
	For sake of comparison, in the plots (a) and (c), a curve concerning a given value of $\eta$
	corresponds to the curve related to the same value of $\kappa/G$ in plot (b).}
\end{figure}

The probability of finding one excitation in mode $M_{2}$ if it is initially
in this mode is
\begin{equation}
p\left( t\right) =\left| \overset{N}{\underset{j=1}{\sum }}%
P_{2,j}^{2}e^{-i\lambda _{j}t}\right| ^{2}\text{.}
\end{equation}
In Ref. \cite{art18}, two regimes were found for the dynamics of such a probability
calculated by using the master equation corresponding to Hamiltonian (\ref
{H1}) with resonant modes $M_1$ and $M_2$:
\begin{eqnarray}
\label{master}
\frac{d}{dt}\hat{\rho}&=&\kappa\left(2\hat{a}_{1}\hat{\rho}\hat{a}_{1}^{\dagger}
-\hat{\rho}\hat{a}_{1}^{\dagger }\hat{a}_{1}-\hat{a}_{1}^{\dagger }\hat{a}_{1}\hat{\rho}\right)\\
&-&i\left(\Omega\left[\hat{a}_{1}^{\dagger }\hat{a}_{1}+\hat{a}_{2}^{\dagger }\hat{a}_{2},\hat{\rho}\right]
+G\left[\hat{a}_{2}^{\dagger }\hat{a}_{1}+\hat{a}_{1}^{\dagger }\hat{a}_{2},\hat{\rho}\right]\right).\notag
\end{eqnarray}
Here, the density operator $\hat{\rho}$ stands for the state of the composed system $M_1+M_2$.
In one of the regimes, the increasing of the dissipation constant $\kappa$ of mode
$M_{1}$ leads to the decreasing of the permanence of the excitation in mode $%
M_{2}$. This is expected, since mode $M_{2}$ is connected to the environment
only through mode $M_{1}$. The other regime was called \emph{Zeno regime}:
there, the increasing of the interaction of $M_{1}$ with the environment
inhibits the transition of the excitation from $M_{2}$ to $M_{1}$, leading
to the enhancement of the probability of finding the excitation in $M_{2}$.
The turning point between these regimes occurs for $\kappa=2G$, where $G$ is the
unitary coupling constant between the modes of interest. In the Appendix, we
show that such a turning point corresponds to $\Gamma =\sqrt{2\Delta G/\pi }$,
what is corroborated by Figs. (\ref{fig1a}) and (\ref{fig2a}). The occurrence of
two regimes may be understood with the help of
the following analytically calculable cases: for $\Gamma =0$, $p\left(
t\right) =\left( 1+\cos \left( 2Gt\right) \right) /2$; for $G=0$, $p\left(
t\right) =1$. Depending on the relation between $\Gamma $ and $G$, the
dynamics can
be approached to one of these limiting cases.
By comparing Figs. (\ref{fig1a}) and (\ref{fig2a}) with Figs. (\ref{fig1b}) and (\ref{fig2b}),
respectively, we see that the Hamiltonian and the master equation approaches exhibit good agreement.
In order to achieve such an agreement, we have to pay attention to two aspects concerning the
parameter $\Delta$: it must be large enough so that environmental modes with frequencies
out of the interval $\left[ \Omega - \Delta / 2 , \Omega + \Delta / 2 \right]$ have negligible
action on the system (as pointed out in Ref. \cite{art19}, the relevant environmental modes
are the ones with frequencies around
the frequencies of the normal modes of the system $\Omega \pm G$); the ratio
$\Delta / \left( N-2 \right)$ must be small, allowing
the use of the limit of dense spectrum.

\section{Sequence of Measurements}

In this section we study the dynamics of two resonant coupled modes
($M_{1}$, $M_{2}$) and $N$ atoms interacting, one at the time,
with mode $M_{1}$. The sequence of interacting atoms represents,
in the limit of instantaneous interactions, a continuous measurement
of the $M_{1}$ excitation number. The investigation shows that the two
regimes reported in Ref. \cite{art18} and in the previous section are also present
if a sequence of atomic interactions is performed on mode $M_{1}$. These
results illustrate the relation between continuous measurement of a quantum
system and the dynamics governed by a master equation.

The results are obtained by numerical simulations, where we consider, as in Ref. \cite{art18},
the system of modes $M_{1}$ and $M_{2}$ in the initial state $|0_{1},1_{2}\rangle$.
A sequence of two level atoms interacts with mode $M_{1}$. The atoms are prepared in
the ground
state $|g\rangle$ and interact with mode $M_{1}$ one at the time. The Hamiltonian of
the global system for the interaction of the $k$-th atom is given by
\begin{equation}
\hat{H}^{(k)}= \Omega ( \hat{a}_{1}^{\dagger }\hat{a}_{1}+ \hat{a}_{2}^{\dagger }\hat{a}_{2} )+
G(\hat{a}_{1}^{\dagger }\hat{a}_{2}+\hat{a}_{2}^{\dagger }\hat{a}_{1})+\frac{\Omega}{2}\hat{\sigma}_{z}^{(k)}+ g(\hat{a}_{1}^{\dagger}\hat{\sigma}_{-}^{(k)}+\hat{a}_{1}\hat{\sigma}_{+}^{(k)}), \label{Hamiltonian}
\end{equation}
where $\hat{a}_{1}^{\dagger }$ ($\hat{a}_{1}$) and $\hat{a}_{2}^{\dagger }$ ($\hat{a}_{2}$) are creation
(annihilation) operators for modes $M_{1}$ and $M_{2}$, $\Omega $
their frequency, $G$ the modes coupling constant, $\hat{\sigma}_{z}^{(k)}=|e^{(k)}\rangle\langle e^{(k)}|-|g^{(k)}\rangle\langle g^{(k)}|$, $\hat{\sigma}_{-}^{(k)}=|g^{(k)}\rangle\langle e^{(k)}|$, $\hat{\sigma}_{+}^{(k)}=|e^{(k)}\rangle\langle g^{(k)}|$ and $g$ the coupling constant for the interaction between $k$-th atom and mode $M_{1}$. Here, $\left|g^{(k)} \right\rangle$ and $\left|e^{(k)}\right\rangle$
stand for the ground and the excited states of the $k$-th atom, respectively.

After each interaction we perform the trace over the corresponding
atomic system, i.e., we do not consider the final state of the atoms.
We also assume that the coupling constant $g$ scales as $1/\sqrt{t_{int}}$,
where $t_{int}$ is the interaction time of each atom \cite{art20}. The overall effect
of these atomic interactions with a cavity mode is a dissipative evolution of
the mode. The effective dissipative constant related to this process is $\kappa=g^{2}t_{int}$.

In Figs. (\ref{fig1c}) and (\ref{fig2c}) we shown two regimes for the probability of finding the excitation
in mode $M_{2}$ ($\mathcal{P}(t)$). In the dissipative regime ($\kappa<2G$), the increasing
of the dissipative constant $\kappa$ leads to the decreasing of $\mathcal{P}(t)$. In the second
regime ($\kappa>2G$), the increasing of the dissipative constant $\kappa$ leads to the increasing
of $\mathcal{P}(t)$, preserving then the excitation in mode $M_{2}$. It is worth to note
that the agreement with the master equation results is reached in the limit
of vanishing interaction time, $t_{int}$. The increasing of interaction time
leads the curves away from the ones obtained by the master equation.

\section{Conclusion}

In the present work we investigate the two regimes of the system of two
coupled modes, induced by two completely different dynamics. In the first
one we consider the mode $M_{1}$ linearly coupled to a finite number of
harmonic oscillators, and in the second one we consider such mode interacting
with $N$ atoms, one at the time. Both dynamics, in appropriate limits,
can describe the regimes obtained in \cite{art18} using a master equation.
In the first model,  when the number of harmonic oscillators goes to
infinity the coupling between them and the system of interest can
simulate the interaction with the environment. In the second model, the
interaction with $N$ atoms, when the interaction time goes to zero,
simulates continuous measurements. Therefore, the present results
illustrate the idea that the interaction between system of interest and
environment can be interpreted as continuous measurement on the system of
interest. The results for both dynamics were obtained using numerical
simulations. As no approximations were used in the calculations, the
analysis of intermediate scenarios, out of the master equations limits, is possible.

\section{Acknowledgements}
This work was partially supported by the brazilian agencies CNPq
and FAPEMIG.

\appendix
\section{Establishing the relation between the effective dissipation constant and
the parameters used in the Hamiltonian approach}

The master equation employed in Ref. \cite{art18} is
given by Eq. \eqref{master}.
It describes the dynamics of two linearly coupled resonant modes, one of
them interacting with an environment at zero absolute temperature. Such a
master equation may be derived as an approximation of the dynamics emerging
from the Hamiltonian $\hat{H}$, leading, under the specifications in Eq. (%
\ref{H2}), to
\begin{equation}
\kappa=\frac{\Gamma ^{2}}{2\left( N-2\right) }\overset{\tau _{c}}{\underset{-\tau
_{c}}{\int }}d\tau \left\{ \overset{\frac{N-2}{2}}{\underset{j=1}{\sum }}%
\left[ e^{i\left( \nu _{j}-G\right) \tau }+e^{i\left( \nu _{j}+G\right) \tau
}\right] \right\} \text{,}
\end{equation}
where $\nu _{j}=j\Delta /\left( N-2\right) $ and $\tau _{c}$ is a value of $%
\left| \tau \right| $ beyond which the summations above are negligible \cite{art19}.
For $N$ sufficiently large, we can take the limit of
dense spectrum, resulting in
\begin{equation}
\kappa=\frac{\Gamma ^{2}}{2\Delta }\overset{\frac{\Delta }{2}}{\underset{\frac{%
\Delta }{N-2}}{\int }}d\omega \overset{\tau _{c}}{\underset{-\tau _{c}}{\int
}}d\tau \left[ e^{i\left( \omega -G\right) \tau }+e^{i\left( \omega +G\right) \tau } \right]%
\text{.}
\end{equation}
Since the integrand is assumed to be negligible for $\left| \tau \right|
>\tau _{c}$, we change $\pm \tau _{c}$ for $\pm \infty $, which leads to
\begin{equation}
\kappa=\frac{\pi \Gamma ^{2}}{\Delta }\overset{\frac{\Delta }{2}}{\underset{\frac{%
\Delta }{N-2}}{\int }}d\omega \left[ \delta \left( \omega -G\right) +\delta
\left( \omega +G\right) \right] \text{.}
\end{equation}
Observing that the term related to $\delta \left( \omega +G\right) $
vanishes, we find, for $\Delta /\left( N-2\right) <G<\Delta /2$,
\begin{equation*}
\kappa=\frac{\pi \Gamma ^{2}}{\Delta }\text{.}
\end{equation*}
The transition to the Zeno regime were found in Ref. \cite{art18} for $\kappa=2G$. This
corresponds to
\begin{equation}
\Gamma =\sqrt{\frac{2\Delta G}{\pi }}\text{.}
\end{equation}

\newpage

\newpage

\end{document}